\documentclass[conference]{IEEEtran}
\IEEEoverridecommandlockouts
\usepackage{cite}
\usepackage{amsmath,amssymb,amsfonts}
\usepackage{mathtools}
\usepackage{algorithmic}
\usepackage{algorithm}
\usepackage{graphicx}
\usepackage{textcomp}
\usepackage{xcolor}
\usepackage{array}
\usepackage[caption=false]{subfig}
\usepackage{stfloats}
\usepackage{url}
\usepackage{verbatim}

\usepackage{tabularx}

\usepackage{todonotes}

\usepackage{amsthm}
\usepackage{bm}

\def\BibTeX{{\rm B\kern-.05em{\sc i\kern-.025em b}\kern-.08em
    T\kern-.1667em\lower.7ex\hbox{E}\kern-.125emX}}

\newtheorem{theorem}{Theorem}

\theoremstyle{definition}

\theoremstyle{remark}
\newtheorem*{remark}{Remark}

\newcommand{\bd}{\mathbf{d}}

\newcommand{\br}{\mathbf{r}}

\newcommand{\bTheta}{\bm{\Theta}}

\newcommand{\diff}{\mathrm{d}}

\DeclareMathOperator{\sinc}{sinc}
\DeclareMathOperator{\laplacian}{\nabla^2}
\DeclareMathOperator{\e}{e}

\DeclareMathOperator{\opherglotz}{\mathcal{H}\!}

\newcommand{\trans}{\mathsf{T}}

\newcommand{\opift}{\mathcal{F}^{-1}}
\newcommand{\kernel}{K}
\DeclarePairedDelimiter\ceil{\lceil}{\rceil}
\DeclarePairedDelimiter\floor{\lfloor}{\rfloor}

\newcommand{\domain}{\Omega}
\newcommand{\setreal}{\mathbb{R}}

\newcommand{\setsensing}{\mathcal{M}}
\newcommand{\setsphere}{\mathcal{S}}

\newcommand{\lightspeed}{c}

\newcommand{\wavenumber}{\kappa}
\newcommand{\wavevector}{\mathbf{k}}
\newcommand{\wavelength}{\beta}
\newcommand{\sfield}{u}
\newcommand{\sbasis}{\psi}

\newcommand{\sdensity}{\sigma}

\newcommand{\eigval}{\lambda}

\newcommand{\aperture}{L}
\newcommand{\spos}{r}
\newcommand{\dof}{\mathrm{DoF}}
\newcommand{\dofth}{\dof_{\mathrm{th}}}
\newcommand{\dofnumA}{\dof_{90\%}}
\newcommand{\dofnumB}{\dof_{99\%}}
\newcommand{\vangle}{\bTheta}

\newcommand{\vdir}{\bd}
\newcommand{\vpos}{\br}

\begin{document}

\title{A Universal Framework for\\ Holographic MIMO Sensing
}

\author{Charles Vanwynsberghe$^{1}$, Jiguang He$^1$, Mérouane Debbah$^2$\\
    $^1$Technology Innovation Institute, 9639 Masdar City, Abu Dhabi, UAE\\
    $^2$Khalifa University of Science and Technology, PO Box 127788, Abu Dhabi, UAE
}

% \and
% \and
% \and
% \and
% \and

\maketitle

\begin{abstract}
This paper addresses the sensing space identification of arbitrarily shaped continuous antennas. In the context of holographic multiple-input multiple-output (MIMO), a.k.a. large intelligent surfaces, these antennas offer benefits such as super-directivity and near-field operability. The sensing space reveals two key aspects: (a) its dimension specifies the maximally achievable spatial degrees of freedom (DoFs), and (b) the finite basis spanning this space accurately describes the sampled field. Earlier studies focus on specific geometries, bringing forth the need for extendable analysis to real-world conformal antennas. Thus, we introduce a universal framework to determine the antenna sensing space, regardless of its shape. The findings underscore both spatial and spectral concentration of sampled fields to define a generic eigenvalue problem of Slepian concentration. Results show that this approach precisely estimates the DoFs of well-known geometries, and verify its flexible extension to conformal antennas.

\end{abstract}


\section{Introduction}
The core principle of holographic multiple-input multiple-output (MIMO) is based on the concept of packing elements more densely than what the Nyquist criterion requires to build an antenna. Also referred to as large intelligent surface (LIS) \cite{Dardari2020}, such an antenna is ultimately thought as a continuous object when element packing is extremely dense, resulting in infinitesimal inter-element spacing. Higher directivity, signal-to-noise ratio, and near-field communication performance are expected from the deployment of LISs \cite{An2023a}.

Holographic MIMO communication is fundamentally characterized by the continuous analogue of the singular value decomposition of the channel propagation model between a pair of transmitting and receiving LISs \cite{Miller2000}. Two important aspects stem from it. First, the highest number of independent data streams between the two LISs, named \textit{spatial degrees of freedom} (DoFs), dictates the maximal multiplexing gain of MIMO communications. The DoFs correspond to the number of non-degenerate singular values of the decomposition \cite{Dardari2020}. Despite dense packing, physical-based limitation exists as the DoFs do not simply scale with the number of elements in the LIS. Second, MIMO communication performance is optimal when the transmitted (resp. received) data streams are multiplied by the right (resp. left) singular functions\footnote{With continuous analogue of the singular value decomposition \cite{Townsend2015}, we refer to (left and right) singular \textit{functions} rather than singular \textit{vectors}, for the well-justified reason that the resulting basis is infinite-dimensional.} of the decomposition.

Performance analysis based on this singular value decomposition has a caveat, as it evidences the performance of the modeled channel only. In this paper, we rather focus on the \textit{intrinsic} performance of the receiving antenna, irrespective of the propagation channel that could be involved. Earlier works based on sampling theory follow that logic, to extract the maximal reachable DoFs of linear, square and cubic LISs \cite{Pizzo2020,Pizzo2020a}. They also show how the proper choice of functions from the Fourier basis offers a fair representation of the sensing space for these particular geometries, when the aperture is large with respect to the wavelength. Similar works focus on the ball or disk geometries \cite{Kennedy2007,Poon2005}. However, they cannot be applied universally to LISs of arbitrary geometry, whereas real-world applications may necessitate the use of a conformal antenna that fits the shape of its supporting structure \cite{Josefsson2014}.

In this paper, we propose a universal framework to identify the sensing space structure of a continuous LIS. By \textit{universal} we mean that:
\begin{enumerate}
    \item the LIS shape is arbitrary, and
    \item the identified sensing space is valid provided that the propagation media is homogeneous and isotropic in a star-shaped volume that contains the LIS, but the propagation conditions are arbitrary outside that volume.
\end{enumerate}
We focus on deterministic scalar electric fields. Starting from the homogeneous Helmholtz equation, the approach is founded on the fact that any Helmholtz solution can be described by an equivalent decomposition into plane waves \cite{Moiola2011,Colton2001}. This plays a central role in this paper: it implies that sensed electric fields have a spectral support, in the wavevector domain, limited to a sphere with a radius equal to the wavenumber. Considering also that the sampling operation by the LIS is also volume-limited in the spatial domain, we show how these properties can be reformulated into an eigenvalue problem of Slepian concentration in integral form \cite{Simons2010,Slepian1964}. Numerical simulations of this paper show that
\begin{itemize}
\item by extension from the one-dimensional case \cite{Slepian1961}, the non-degenerate eigenvalues of this problem also provide the maximal DoFs that can be achieved,
\item the space spanned by their related eigenfunctions (also named Slepian functions\footnote{Slepian functions are also referred to as prolate spheroidal functions; in this paper we opt for the former name.}) offers an accurate approximation of the sampled signals.
\end{itemize}

The rest of the paper is organized as follows. Sec. \ref{sec:model_and_approx} describes the scalar electric field by its propagation model, and its general approximation by plane waves decomposition from the literature. Sec. \ref{sec:universal_subspace} exposes the eigenvalue problem of Slepian concentration to identify the structure of the universal sensing space. Finally, sec. \ref{sec:numerical_analysis} provides a numerical analysis: first we verify that the numerical DoFs and sensing space corroborate with well-known cases from \cite{Pizzo2020,Poon2005}, and then the asset of the proposed approach is illustrated with a paraboloid conformal antenna.

\section{Propagation Model and Its Approximated Solutions}
\label{sec:model_and_approx}

\subsection{Propagation of Electric Field in Homogeneous Media}

We consider the propagation of the scalar electric field $\sfield$ in a bounded domain $\domain$ of the three-dimensional space. The waves propagate at the speed $\lightspeed$ in this domain, assumed to be homogeneous and isotropic. The propagation is described at time $t$ and space $\vpos = [\spos_1, \spos_2, \spos_3] \in \domain$ by the wave equation
\begin{equation}
    \label{eq:eqdiffwave}
    \laplacian \sfield(\vpos, t) - \dfrac{1}{\lightspeed^2} \frac{\partial^2}{\partial t^2}  \sfield(\vpos, t) = 0 \, ,
\end{equation}
where $\laplacian = \sum_{n = 1}^3 \frac{\partial^2}{\partial r_n^2} $ is the Laplacian operator. The wave equation remains general for fields with arbitrary frequency content, but can be rewritten for the case of narrowband field at frequency  $\omega$ in radians/second by considering harmonic solutions of the form $\sfield(\vpos, t) = \sfield(\vpos) e^{\jmath \omega t}$. It leads to the Helmholtz equation
\begin{equation}
        \label{eq:eqdiffhelmholtz}
    \nabla \sfield(\vpos) + \wavenumber^2 \sfield(\vpos) = 0 \, ,
\end{equation}
where $\wavenumber = 2\pi / \wavelength$ is the wavenumber, and $\wavelength = 2\pi \lightspeed / \omega$ is the wavelength.

A general solution of the homogeneous equation is, up to a constant,
\begin{equation}
    \label{eq:solplanewave}
    \sfield(\vpos) = e^{\jmath \wavenumber \vdir(\vangle) . \vpos}
\end{equation}
with $\vangle = [\theta, \phi]^\trans$ being the angles from the polar coordinate system, and $\vdir(\vangle) = [\sin\theta\cos\phi, \sin\theta\sin\phi, \cos\theta]^\trans$ the unit-norm vector pointing towards $\vangle$. Solution \eqref{eq:solplanewave} describes a plane wave propagating towards $\vdir(\vangle)$. By extension, any function derived from a continuous expansion of plane waves from the sphere also satisfies the Eq. \eqref{eq:eqdiffhelmholtz}. Such functions, known as \textit{Herglotz wave functions}, they are of the form:
\begin{equation}
    \label{eq:helglotzwvfunc}
     \sfield(\vpos) = \opherglotz \sdensity  = \iint_{4\pi} \sdensity(\vangle)  e^{\jmath \wavenumber \vdir(\vangle) . \vpos} \diff\vangle
\end{equation}
with $ \diff \vangle = \sin(\theta) \diff \theta \diff \phi$ the differential solid angle, and $\sdensity(\vangle)$ the wave density. The function $\opherglotz$ is defined when $\sdensity(\vangle)$ is square integrable on the sphere, \textit{i.e.} when the wave density has finite energy.

\subsection{Approximation of Helmholtz Solutions by Herglotz Wave Functions}

As Herglotz wave functions form a subset of the solutions to the differential equation \eqref{eq:eqdiffhelmholtz}, a natural question arise: \textit{can any solution be approximated by a Herglotz function with a corresponding wave density $\sdensity(\vangle)$?} Earlier studies claim so by invoking the density property of polynomial functions in Sobolev spaces \cite{Colton2001,Weck2004}, provided that the domain $\domain \subset \setreal^3$ is star-shaped \cite{Melenk1999}. Although functional analysis goes beyond the scope of this paper, some details are provided below to grasp the relation between Eq. \eqref{eq:eqdiffhelmholtz} and the function $\opherglotz$ . Let the $H_1$ norm of $\sfield$ be defined on $\domain$ as:
\begin{equation}
    \label{eq:normh1}
    \|\sfield \|_{H_1}^2 = \iiint_{\domain} |\sfield (\vpos)|^2 + \sum_{n=1}^{3} \left|\dfrac{\partial \sfield (\vpos)}{\partial \vpos_n}\right|^2  \diff \vpos \,.
\end{equation}
The virtue of such a norm is to establish a metric between two functions, both in terms of scalar gradient values. Colton \textit{et al.} develop the following theorem.
\begin{theorem}[Th. 2.3 \cite{Colton2001}]
\label{th:theorem_density}
Assume $\sdensity$ to be square integrable on the sphere. Then, the set of Herglotz wave functions is dense in the space of solutions to the Helmholtz equation in the volume $\domain$, with respect to the $H_1$ norm.
\end{theorem}
In other words, for any given solution $\sfield (\vpos)$ with $\vpos \in \domain$, there exists a wave density $\sdensity$ such that the metric $\| \sfield - \opherglotz\sdensity\|_{H_1}^2$ vanishes. The approximation holds for both the electric field and its gradient, although the term $\sfield$ usually plays the most important role in the context of wireless communication and array processing, as we mostly rely on measurements of $\sfield$.\footnote{We assume that the field is measured by conventional scalar sensors other than a vector-sensor array~\cite{Nehorai1994}.}

\begin{remark}
For sake of simplicity, the paper focuses on scalar electric fields. However, Colton \textit{et al.} extended \textbf{Theorem \ref{th:theorem_density}} by starting from the system of Maxwell equations \cite{Colton2001a}, and showed that the vectorized form of Herglotz functions approximate electric vector fields. Without loss of generality, the presented results can be generalized to encompass wireless communication scenarios including polarization.
\end{remark}

\textbf{Theorem \ref{th:theorem_density}} is a fundamental milestone to represent any scalar field as an ``infinite" sum of plane waves. Nevertheless, the abstract concept of set density does not indicate anything about the approximation error in practice. On the other hand, Moiola \textit{et al.} \cite{Moiola2011} successfully derive an upper bound of this error, even when the number of plane waves is finite. If $P$ vectors $\vdir$ are regularly distributed over the unit sphere, they show that a set of complex coefficients $\alpha_1, \alpha_2, \dots, \alpha_P$ exist such that
\begin{equation}
    \label{eq:uapproxsum}
    \sfield(\vpos) \approx \sum_{n=1}^P \alpha_n e^{\jmath \wavenumber \vpos . \vdir(\vangle_n)},
\end{equation}
provided that $P$ is sufficiently large, and $\domain$ is star-shaped. In the three-dimensional case, the approximation error decays exponentially with respect to $P$ \cite[Corollary 5.5]{Moiola2011}. This statement is stronger than \textbf{Theorem \ref{th:theorem_density}}, as the bound guarantees that the error vanishes rapidly.

\subsection{Take-Home Message}
Approximations in Eqs. \eqref{eq:helglotzwvfunc} and \eqref{eq:uapproxsum} provide a generalized way to describe the electric field in a homogeneous and isotropic volume. Unlike the reported works from \cite{Ikehata2012,Koyama2019,Mignot2013} dealing with inverse problems, we do not focus on the values that $\sdensity(\vangle)$'s or $\alpha_n$'s take to reconstruct the field. In the current context, the central role of these approximations is that the set of plane waves $\left\{ e^{\jmath \wavenumber \vpos.\vdir(\vangle)} | \vangle \in [0, \pi] \times [0, 2\pi], \vpos \in \domain  \right\}$ spans the space containing all scalar electric fields in $\domain$.

\begin{figure}
    \centering
     \includegraphics[]{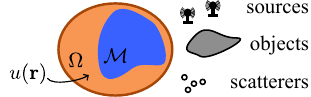}
    \caption{Considered scenario: a LIS of volume $\setsensing$ samples the electric field in the homogeneous and isotropic volume $\domain$. There is no restriction outside $\domain$ in the presence of non-linearities, scattering, diffraction or refraction.}
    \label{fig:scheme_rx_only}
\end{figure}

Suppose then that a LIS is contained inside $\domain$, with no additional restriction on its geometry  -- see Fig. \ref{fig:scheme_rx_only}. Then the sensing space in which its noiseless measurements lie also falls into the space spanned by plane waves. Interestingly, the required conditions do not impose any restriction \textit{outside} $\domain$. It offers a high flexibility in the possible scenarios to consider: the presence of non-linearities or objects leading to reflection, diffraction, refraction or scattering outside $\domain$ does not alter the sensing space structure of the LIS located inside $\domain$. Based on these elements, the next section introduces a general approach to obtain the structure of the sensing space of a LIS of finite aperture.

\section{Universal Subspace of Deterministic Fields}
\label{sec:universal_subspace}

From a signal processing perspective, this section shows how to identify the space that electrical fields sensed by a given LIS occupy, the LIS volume being defined by $\setsensing \subset \domain$ -- cf Fig. \ref{fig:scheme_rx_only}.
To do so, the approach consists in finding a set of orthogonal functions $\sbasis_1, \sbasis_2, \dots, \sbasis_N$ such that $\sfield(\vpos)$ can be regarded as a linear combination of them for $\vpos \in \setsensing$, that is:
\begin{equation}
    \label{eq:slepiandecomposition}
\sfield(\vpos) \approx \sum_{i=1}^{N} \gamma_i \sbasis_i(\vpos), ~ \vpos \in \setsensing
\end{equation}
with $\gamma_i = \iiint_\setsensing \sfield(\vpos) \sbasis_i^*(\vpos) \diff\vpos$ complex coefficients.
Two properties of the sampled field are highlighted and leveraged in the following.

First, Sec. \ref{sec:model_and_approx} reveals the spectral structure of $\sfield$ in the wavevector domain. With $\wavevector \in \setreal^3$ being the wavevector, and $\hat \sfield$ being the spectrum of $\sfield$, the inverse Fourier transform (IFT) $\opift$ in the volume
\begin{equation}
    \label{eq:ift}
    \sfield(\vpos) = \opift \hat \sfield =  \dfrac{1}{(2\pi)^{3}} \iiint_{\setreal^3} \hat \sfield(\wavevector) e^{\jmath \wavevector . \vpos} d\wavevector
\end{equation}
has a direct relation with the Herglotz wave function. Indeed, by changing the integral of Eq. \eqref{eq:ift} from Cartesian to spherical coordinate system one can evidence that
\begin{equation}
    \opherglotz \{\sdensity(\vangle)\} = \frac{(2\pi)^3}{\wavenumber^2} \opift \{ \delta(\|\wavevector\| - \wavenumber) \sdensity(\vangle)\}
\end{equation}
with $\delta$ being the Delta Dirac function.
Therefore, fields are bandlimited in the wavevector domain on the sphere $\setsphere$ of radius $\wavenumber$:
\begin{equation}
    \setsphere = \{\wavevector \in \setreal^3, \| \wavevector \| = \wavenumber \}.
\end{equation}

Second, the sampling operation in the volume $\setsensing$ can be regarded as capturing a signal whose value equals $\sfield(\vpos)$ for $\vpos \in \setsensing$, and zero elsewhere. The signal is thus also limited in the spatial domain.

\subsection{Slepian Functions of Sampled Fields in the Volume}

The representation of signals that are limited in both spatial and wavenumber domains belongs to a class of problems that were addressed first by Slepian, Landau and Pollak \cite{Slepian1961} for one-dimensional signals. Here, we leverage the approach in the Cartesian volume to derive a basis of functions $\sbasis_i$'s -- also named Slepian functions.

Slepian's concentration problem can be formulated as follows: we search the function $\sbasis$ which maximizes the concentration of its spectrum $\hat\sbasis$ in $\setsphere$, under the constraint that its support is limited in $\setsensing$:
\begin{equation}
    \label{pb:variational__slepian_problem}
    \begin{split}
    \eigval = \max_{\sbasis = \opift{\hat \sbasis}} & \dfrac{\iiint_{\setsphere} | \hat \sbasis(\wavevector)|^2 \diff\wavevector}{\iiint_{\setreal^3} | \hat \sbasis(\wavevector)|^2 \diff\wavevector} \\
     \mathrm{s.t.~}& \sbasis(\vpos) = 0 \mathrm{~for~} \vpos \in \setreal^3 \setminus \setsensing .
    \end{split}
\end{equation}
Sampled signals satisfying the variational problem \eqref{pb:variational__slepian_problem} in three dimensions are solutions of the Fredholm integral equation \cite{Slepian1964, Simons2010}:
\begin{equation}
    \label{eq:eigenvalue_problem}
    \iiint_{\setsensing} \kernel(\vpos, \vpos') \sbasis(\vpos') \diff \vpos' = \eigval \sbasis(\vpos) , \, \vpos \in \setsensing
\end{equation}
with the kernel function $K(\vpos, \vpos')$ in the form of
\begin{equation}
    \label{eq:kernel_slepian}
    K(\vpos, \vpos') = (2\pi)^{-3} \iiint_{\setsphere} \e^{\jmath \wavevector . (\vpos - \vpos')} \diff\wavevector.
\end{equation}
Eq. \eqref{eq:eigenvalue_problem} is an eigenvalue problem, admitting an infinite collection of solutions of eigenvalues $\eigval_1 > \eigval_2 > \dots > 0$ and their related orthogonal eigenfunctions $\sbasis_1, \sbasis_2, \dots$ (named Slepian functions) of unitary energy.

Note that the variational problem \eqref{pb:variational__slepian_problem} could be formulated as \textit{searching $\hat \sbasis$ maximizing the concentration of its IFT $\sbasis$ in the volume $\setsensing$, under the constraint that its support is limited in $\setsphere$}. By permuting the maximization criterion and the constraint, the related eigenvalue equation \eqref{eq:eigenvalue_problem} turns out to become a Fredholm integral in the wavevector domain, and the kernel \eqref{eq:kernel_slepian} is changed to an integral in the sampling domain.
%Both of their solutions coincide \cite{Slepian1961}.
We favor the first, more convenient and general approach here, as the integral \eqref{eq:kernel_slepian} is invariant from the sampling volume $\setsensing$.

In spherical coordinates, with $\diff \wavevector = \|\wavevector\|^2 \diff \|\wavevector\| \diff\vangle$ and $\diff\vangle = \sin\theta\diff\theta\diff\phi$, the integral \eqref{eq:kernel_slepian} becomes
\begin{equation}
\begin{split}
    & K(\vpos, \vpos') =  (2\pi)^{-3} \iiint \delta(\|\wavevector\| - \wavenumber) \e^{\jmath \wavevector . (\vpos - \vpos')} \|\wavevector\|^2 \diff \|\wavevector\| \diff\vangle \\
    &= (2\pi)^{-3} \int_{\setreal_+} \delta(\|\wavevector\| - \wavenumber) \|\wavevector\|^2 \diff \|\wavevector\| \iint_{4\pi} \e^{\jmath \wavenumber \vdir(\vangle) . (\vpos - \vpos')} \diff\vangle.
\end{split}
\end{equation}
The radial integral on the wavevector norm reduces to $\int_{\setreal_+} \delta(\|\wavevector\| - \wavenumber) \|\wavevector\|^2 \diff \|\wavevector\| = \wavenumber^2$. Without loss of generality, we simplify the integral by rotating the coordinate system such that $\vpos - \vpos' = [0, 0, \| \vpos - \vpos' \|]$, which yields:
\begin{equation}
\begin{split}
    \iint_{4\pi} \e^{\jmath \wavenumber \vdir(\vangle) . \|\vpos - \vpos'\|} \diff\vangle
    &= \int_{2\pi} \diff\phi \int_{\pi} \e^{\jmath \wavenumber \|\vpos - \vpos'\| \cos\theta} \sin(\theta) \diff\theta \\
    &= 2\pi \left[ - \frac{1}{\jmath \wavenumber \|\vpos - \vpos'\|} \e^{\jmath \wavenumber \|\vpos - \vpos'\| \cos\theta} \right]_{0}^{\pi} \\
    &= 2\pi . 2 \frac{\sin(\wavenumber \|\vpos - \vpos'\|)}{\wavenumber \|\vpos - \vpos'\|}.
\end{split}
\end{equation}
Combining the integral terms together finally gives
\begin{equation}
\begin{split}
    \label{eq:kernel_helmoltz}
    K(\vpos, \vpos')
    &= (2\pi)^{-3} \wavenumber^2  4\pi \sinc(\wavenumber  \|\vpos - \vpos'\|) \\
    &= \frac{2}{\wavelength^2} \sinc(\wavenumber  \|\vpos - \vpos'\|).
\end{split}
\end{equation}
where $\sinc(x) = \sin(x)/x$.

\subsection{Connection with Bandlimited Signals on the Line, and Degrees of Freedom}

With the kernel \eqref{eq:kernel_helmoltz}, Eq. \eqref{eq:eigenvalue_problem} describes a spatial-domain convolutional integral. Interestingly, this problem for three-dimensional scalar fields is closely related to the one for one-dimensional signals $g(t)$'s that are bandlimited on the interval $[-W, W]$ and sampled on the segment $[-T/2, T/2]$, i.e., with $\hat g$ the spectrum of $g$,
\begin{equation}
    g(t) = \frac{1}{2\pi} \int_{-W}^{W} \hat g(\omega) \e^{\jmath \omega t} \diff\omega \text{ and } \hat g(\omega) =  \int_{-\frac{T}{2}}^{\frac{T}{2}} g(t) \e^{-\jmath \omega t} \diff\omega .
\end{equation}
The related Fredholm integral becomes in that case the time-domain convolution\footnote{Eigenvalue is noted here $\eigval'$ to differentiate it from $\eigval$ in Eq. \eqref{eq:eigenvalue_problem}.} \cite{Slepian1961}:
\begin{equation}
    \label{eq:eigenvalue_problem_line}
    \int_{-T/2}^{T/2} \frac{W}{\pi} \sinc(W(t - t')) g(t') \diff t' = \eigval'  g(t),
\end{equation}
the kernel being derived by the one-dimensional IFT of the segment of length $2W$ and centered at the origin. It turns out to be equal to the kernel \eqref{eq:kernel_helmoltz} -- up to a constant -- when $W=\wavenumber$. This comes from the general fact the IFT of a ball in dimension $n$ equals the IFT of the sphere with same radius in dimension $n+2$ \cite{Vembu1961} -- still up to a constant.

Sorted in the decreasing order, eigenvalues $\eigval_i'$'s from Eq. \eqref{eq:eigenvalue_problem_line} are known to decay at super-exponential rate  \cite{Bonami2021} beyond the threshold $i = \floor{WT/\pi}$, revealing that signals $g(t)$ roughly lie in a subspace of finite dimension. Eigenproblems \eqref{eq:kernel_helmoltz} and \eqref{eq:eigenvalue_problem_line} coincide when field is sampled by a linear LIS. The straightforward consequence is that the signals roughly lie in a space of dimension $\floor{\wavenumber \aperture / \pi}$, where $\aperture$ is the aperture of the linear LIS. In other words, the decomposition Eq. \eqref{eq:slepiandecomposition} holds when $N$ equals -- at least -- this value.

In the context of holographic MIMO, the sensing space dimension indicates the maximal number of independent data streams that the LIS can receive, i.e. its maximal DoFs. By using different approaches, deriving the DoFs analytically remains possible as long as the  geometry remains convenient for calculus \cite{Pizzo2020,Kennedy2007,Poon2005}. However, the generalization is not possible for complex geometries, e.g. for conformal antennas. In that case, solving Eq. \eqref{eq:eigenvalue_problem} becomes helpful to obtain the DoFs numerically, by identifying the index $i$ from which  $\eigval_i$ decays rapidly.

\section{Numerical Analysis of the Sensing Space Structure}
\label{sec:numerical_analysis}

\begin{figure*}
    \subfloat[Scaled eigenvalues $(\eigval_i / \eigval_1)$'s, for different values of $\wavenumber \aperture$. For the white lines: solid is $\dofth$, dashed is $\dof_{90\%}$, and dotted is $\dof_{99\%}$.\label{fig:eigvals_scaled}]{\includegraphics[]{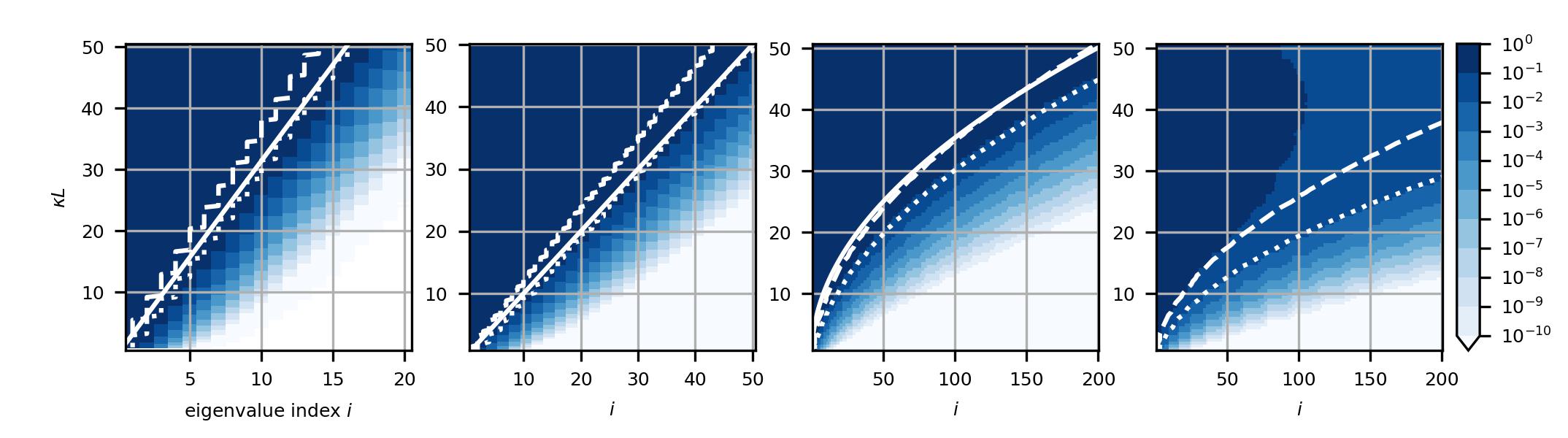}}
    \\
    \includegraphics[trim={0 10mm 0 8mm},clip]{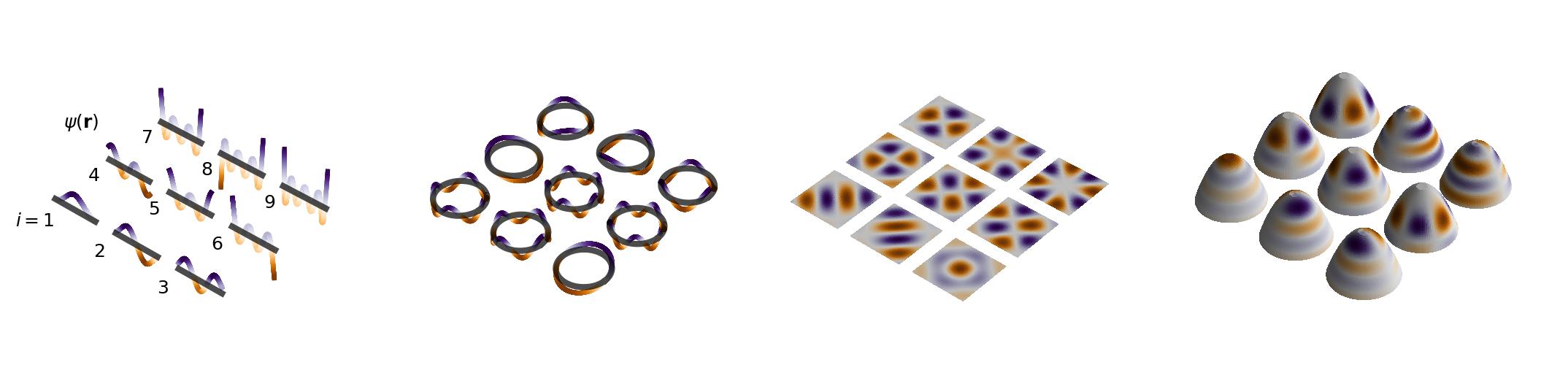}
    \\[-0.1ex]
    \subfloat[Nine first Slepian functions $g_i(\vpos)$'s for $\vpos \in  \setsensing$ (top) and their Fourier transforms $|\hat g_i(\wavevector)|$'s for $\wavevector \in  \setsphere$ (bottom), with $\wavenumber\aperture = 4 \pi$.\label{fig:eigvecs}]{\includegraphics[trim={0 1mm 0 17mm},clip]{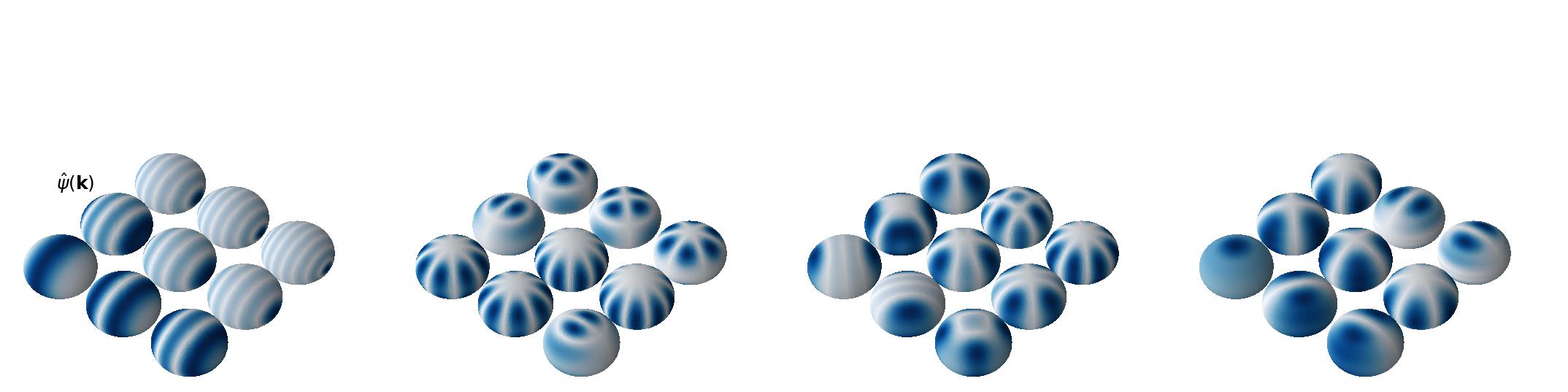}}
    \caption{Signal subspace eigenstructure of four LIS shapes, from left to right: the L-LIS, C-LIS, S-LIS and P-LIS.}
    \label{fig:4antennas}
\end{figure*}

\subsection{Eigenvalue Analysis and Numerical DoFs}

This section provides numerical results for four LIS geometries characterized by an equivalent aperture $\aperture$. First, we verify that numerical DoFs coincide with their analytical counterparts, denoted by $\dofth$. To do so, we choose the linear (L-LIS), circular (C-LIS) and square (S-LIS) cases. Second, we demonstrate the flexibility of solving the Slepian problem with the more complex geometry of a conformal antenna. We choose the paraboloid of revolution (P-LIS), adapted to be mounted on an aircraft nose \cite[chap. 3.3.1]{Josefsson2014}. To the best of our knowledge, $\dofth$ is unknown for the P-LIS. Table \ref{tab:geometries_and_dof} summarizes the geometries and the $\dofth$ of the four LISs.

\renewcommand{\arraystretch}{1.7}
\begin{table}[]
    \caption{The four LISs studied in Fig. \ref{fig:4antennas}.}
    \begin{tabular}{|p{8mm}||p{48mm}|p{19mm}|ll}
        \cline{1-3}
        & \centering Definition of $\setsensing$                                                          & \centering $\dofth$ &  &  \\ \cline{1-3}
        L-LIS & $\spos_1 \in [0, \aperture]$, $r_3 = 0,$ $\spos_2 = 0$                                              & \centering $\frac{\wavenumber\aperture}{\pi}$ \cite{Pizzo2020}  &  &  \\ \cline{1-3}
        C-LIS & $[\spos_1, \spos_2]= 0.5 \,\aperture[\cos\theta, \sin\theta],\hspace*{13mm}$ $\theta \in [0, 2\pi],$ $r_3 = 0$   & \centering $\wavenumber\aperture$ \cite{Poon2005} &  &  \\ \cline{1-3}
        S-LIS & $\spos_1 \in [0, \aperture]$, $\spos_2 \in [0, \aperture]$, $r_3 = 0$                          & \centering $\frac{(\wavenumber\aperture)^2}{4 \pi}$ \cite{Pizzo2020}   &  &  \\ \cline{1-3}
        P-LIS & $[\spos_1, \spos_2] = 0.5\sqrt{\aperture(\aperture - r_3)} [\cos\theta, \sin\theta],$ $\spos_3 \in [0, \aperture]$,  $\theta \in [0, 2\pi]$ & \centering / &  &  \\ \cline{1-3}
    \end{tabular}
    \label{tab:geometries_and_dof}
\end{table}

The vector $\vpos$ is parametized with $r_1$ only for the L-LIS, so the kernel $\kernel(\vpos, \vpos')$ -- reduced to $\kernel(r_1, r_1')$ -- can be regarded as a continuous analogue of a matrix with a real-valued row and column indexes $r_1$ and $r_1'$ \cite{Townsend2015}. The same conclusion holds in the case of the C-LIS, with $\kernel$ indexed by angular parameters. Eigenvalue decomposition of continuous-like matrices are solved up to 15-digit accuracy with the toolbox \textit{chebfun2} \cite{Townsend2015}. For the S-LIS (resp. P-LIS), the integral is discretized with a mesh of 4096 (resp. 4500) cells. The scaled eigenvalues $(\eigval_i / \eigval_1)$'s are plotted for different parameters $\wavenumber \aperture$ in Fig. \ref{fig:eigvals_scaled}. We define $\dofnumA$ (resp. $\dofnumB$) as the minimal number of eigenvalues whose sum accounts for at least $90\%$ (resp. $99\%$) of the full sum. Plots confirm that analytical and numerical DoFs coincide: $\dofth$ is close to $\dofnumA$ for the S-LIS, and close to $\dofnumB$ for the L- and C-LISs. Note that the decay of $\eigval_i$ is slower for surfaces (P- and S-LIS), especially when $\kappa\aperture$ is larger. Because P-LIS is a surface, plots of $\dofnumA$ and $\dofnumB$ exhibit a quadratic trend with respect to $\wavenumber\aperture$.

\subsection{Slepian Functions, and Their Impact on the Wave Density Measurement}

The nine first Slepian functions $\sbasis_i$'s and their Fourier transform magnitudes $|\hat \sbasis_i |$'s on $\setsphere$ are illustrated in Fig. \ref{fig:eigvecs} for the four geometries, with $\wavenumber\aperture = 4\pi$. On the first hand, $\sfield$ can be decomposed as a sum of $\sbasis_i$ as in Eq. \eqref{eq:slepiandecomposition}. The smooth aspect of $\sbasis_i$ scales with the physical parameter $\wavenumber$, since $\sfield$ results from the physical-based propagation. On the other hand,  $|\hat\sbasis_i(\wavevector)|$ restricted to $\wavevector \in \setsphere$ can be interpreted as the radiation pattern of $\sbasis_i$, and reveals the plane wave ``content" of each Slepian function in the wavevector domain. As $\sbasis_i$'s are orthogonal, the Plancherel theorem states that $\hat\sbasis_i$'s are also orthogonal, so the sensed field decomposition in Eq. \eqref{eq:slepiandecomposition} relates to the equivalent decomposition into the wavenumber domain:
\begin{equation}
\hat\sfield(\wavevector) \approx \sum_{i=1}^{N} \gamma_i \hat\sbasis_i(\wavevector).
\end{equation}
In the case where the field is exclusively excited at the pulsation $\omega$, the plots of $|\hat\sbasis_i|$ indicate how the LIS geometry (i.e. $\setsensing$) impacts the sampling resolution of the wave density $\sdensity$ for $N = \floor{\dofth}$. In Fig. \ref{fig:4antennas} (b), the smooth aspect of $\hat\sbasis_i$ scales with both $\wavenumber$ and the geometry aperture.

\begin{figure}
    \includegraphics[]{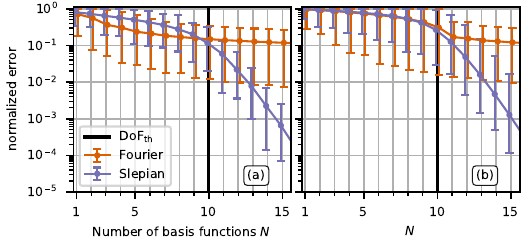}
    \caption{Comparing the expansion of $\sfield(\vpos)$ in the scenario of a LoS channel propagation between two L-LISs, with Fourier \cite{Pizzo2020a} and Slepian basis. Plots are average (dots), minimal and maximal (whiskers) values of normalized error. (a) parallel LISs ($\theta_{\mathrm{Tx}} = \theta_{\mathrm{Rx}} = 0$), (b) randomly rotated LISs, ($0 \leq \theta_{\mathrm{Tx}} , \theta_{\mathrm{Rx}} < 2 \pi$). For (a) and (b): distance $d$ is uniformly sampled between $5$ and $25$ cm.}
    \label{fig:compression_error}
\end{figure}

\subsection{Slepian Functions for Describing Holographic MIMO Channel Propagation}

Finally, this section opens a perspective about the potential of Slepian functions to model holographic MIMO channels. Recent works proved that channels involving a receiving L-LIS are well described by the Fourier plane-wave series expansion \cite[Eq. (43)]{Pizzo2020a} as
\begin{equation}
    \label{eq:fourier_pw_expansion}
\sfield(r_1) \approx \sum_{i = - \floor{N / 2}}^{\ceil{N / 2} - 1} \gamma_i' e^{\jmath 2\pi \frac{r_1}{\aperture}} ~ , r_1 \in [0, \aperture],
\end{equation}
when $N = \floor{\wavenumber\aperture/\pi}$, and $\gamma_i' = \int_{0}^{\aperture} \sfield(r_1)  e^{ - \jmath 2\pi \frac{r_1}{\aperture}} \diff r_1$. Here, we investigate how the expansion given by Slepian functions in Eq. \eqref{eq:slepiandecomposition} is capable of describing channels, in comparison with the model \eqref{eq:fourier_pw_expansion}. To do so, a line of sight (LoS) propagation is simulated between two L-LISs of aperture $\aperture = 5 \wavelength$ in the plane, tilted by angles $\theta_{\mathrm{Tx}}$ and $\theta_{\mathrm{Rx}}$, and whose centers are separated by a distance $d$. The model describing this propagation can be found e.g. in \cite[Eq. (8)]{Miller2000}. In each experiment, the current distribution of the transmitting L-LIS is randomly generated as a smooth polynomial function. With a Rayleigh distance of $12.5$~cm for $\wavelength = 1$ cm, the distance $d$ is uniformly sampled between $5$ and $25$ cm to cover both near and far field LoS propagation.

Results are provided for 2 scenarios: when the LISs remain parallel ($\theta_{\mathrm{Tx}} =\theta_{\mathrm{Rx}} = 0$, cf. Fig. \ref{fig:compression_error} (a)), or are both randomly tilted ($0 \leq \theta_{\mathrm{Tx}} , \theta_{\mathrm{Rx}} < 2\pi$, cf. Fig. \ref{fig:compression_error} (b)) at each experiment. We plot average (dots), minimal and maximal (whiskers) values of normalized errors over 1000 experiments. The normalized error is computed as
\begin{equation}
\frac{\int_{0}^{\aperture} |\sfield(r_1) - \sum_{i=1}^{N} \gamma_i \sbasis_i(r_1)|^2\diff r_1}{\int_{0}^{\aperture} |\sfield(r_1)|^2\diff r_1} \end{equation}
with the Slepian basis, and as
\begin{equation}
    \frac{\int_{0}^{\aperture} |\sfield(r_1) - \sum_{i = -  \floor{N / 2}}^{\ceil{N / 2} - 1} \gamma_i' e^{\jmath 2\pi \frac{r_1}{\aperture}}|^2\diff r_1}{\int_{0}^{\aperture} |\sfield(r_1)|^2\diff r_1}
\end{equation}
with the Fourier basis. Trends show that both Slepian and Fourier basis have an equivalent accuracy for $N = \floor{\dofth}$. Including more functions (i.e. choosing $N>\floor{\dofth}$) provides a better description of $\sfield(r_1)$. Accuracy with the Slepian basis decays faster ($\approx 10^{-3}$ for $N=15$) than with the Fourier basis ($\approx 10^{-1}$ for $N = 15$). As expected from the theory \cite{Pizzo2020a}, increasing $N>\floor{\dofth}$ brings a marginal amelioration to model channels with the Fourier plane-wave series expansion, but Slepian functions seem to be an interesting candidate for a better accuracy.

\section{Conclusion}
We have derived the general Slepian concentration problem of scalar electric fields sampled in the volume. Solving this problem captures the sensing space of a continuous LIS: analysis of eigenvalues gives a fair approximation of the maximal reachable DoFs, and the related Slepian functions provide an accurate basis to represent measured fields. The proposed approach becomes interesting to unlock the performance analysis of complex geometries (e.g. conformal antennas) in the context of holographic MIMO.

\bibliographystyle{IEEEtran}
\bibliography{refs_selected}

\begin{thebibliography}{10}
\providecommand{\url}[1]{#1}
\csname url@samestyle\endcsname
\providecommand{\newblock}{\relax}
\providecommand{\bibinfo}[2]{#2}
\providecommand{\BIBentrySTDinterwordspacing}{\spaceskip=0pt\relax}
\providecommand{\BIBentryALTinterwordstretchfactor}{4}
\providecommand{\BIBentryALTinterwordspacing}{\spaceskip=\fontdimen2\font plus
\BIBentryALTinterwordstretchfactor\fontdimen3\font minus
  \fontdimen4\font\relax}
\providecommand{\BIBforeignlanguage}[2]{{%
\expandafter\ifx\csname l@#1\endcsname\relax
\typeout{** WARNING: IEEEtran.bst: No hyphenation pattern has been}%
\typeout{** loaded for the language `#1'. Using the pattern for}%
\typeout{** the default language instead.}%
\else
\language=\csname l@#1\endcsname
\fi
#2}}
\providecommand{\BIBdecl}{\relax}
\BIBdecl

\bibitem{Dardari2020}
D.~Dardari, ``Communicating with large intelligent surfaces: Fundamental limits
  and models,'' \emph{IEEE Journal on Selected Areas in Communications},
  vol.~38, no.~11, pp. 2526--2537, 2020.

\bibitem{An2023a}
J.~An, C.~Yuen, C.~Huang, M.~Debbah, H.~V. Poor, and L.~Hanzo, ``A tutorial on
  holographic {MIMO} communications—part {II}: Performance analysis and
  holographic beamforming,'' \emph{IEEE Communications Letters}, 2023.

\bibitem{Miller2000}
D.~A.~B. Miller, ``Communicating with waves between volumes: evaluating
  orthogonal spatial channels and limits on coupling strengths,'' \emph{Applied
  Optics}, vol.~39, no.~11, p. 1681, apr 2000.

\bibitem{Townsend2015}
A.~Townsend and L.~N. Trefethen, ``Continuous analogues of matrix
  factorizations,'' \emph{Proceedings of the Royal Society A: Mathematical,
  Physical and Engineering Sciences}, vol. 471, no. 2173, p. 20140585, 2015.

\bibitem{Pizzo2020}
A.~Pizzo, T.~L. Marzetta, and L.~Sanguinetti, ``Degrees of freedom of
  holographic {MIMO} channels,'' in \emph{2020 IEEE 21st International Workshop
  on Signal Processing Advances in Wireless Communications (SPAWC)}.\hskip 1em
  plus 0.5em minus 0.4em\relax IEEE, 2020, pp. 1--5.

\bibitem{Pizzo2020a}
------, ``Spatially-stationary model for holographic {MIMO} small-scale
  fading,'' \emph{IEEE Journal on Selected Areas in Communications}, vol.~38,
  no.~9, pp. 1964--1979, 2020.

\bibitem{Kennedy2007}
R.~A. Kennedy, P.~Sadeghi, T.~D. Abhayapala, and H.~M. Jones, ``Intrinsic
  limits of dimensionality and richness in random multipath fields,''
  \emph{{IEEE} Transactions on Signal Processing}, vol.~55, no.~6, pp.
  2542--2556, jun 2007.

\bibitem{Poon2005}
A.~Poon, R.~Brodersen, and D.~Tse, ``Degrees of freedom in multiple-antenna
  channels: A signal space approach,'' \emph{{IEEE} Transactions on Information
  Theory}, vol.~51, no.~2, pp. 523--536, feb 2005.

\bibitem{Josefsson2014}
\BIBentryALTinterwordspacing
L.~Josefsson and P.~Persson, \emph{Conformal Array Antennas}.\hskip 1em plus
  0.5em minus 0.4em\relax Singapore: Springer Singapore, 2014, pp. 1--35.
  [Online]. Available: \url{https://doi.org/10.1007/978-981-4560-75-7_65-1}
\BIBentrySTDinterwordspacing

\bibitem{Moiola2011}
A.~Moiola, R.~Hiptmair, and I.~Perugia, ``\BIBforeignlanguage{en}{Plane wave
  approximation of homogeneous {Helmholtz} solutions},''
  \emph{\BIBforeignlanguage{en}{Zeitschrift für angewandte Mathematik und
  Physik}}, vol.~62, no.~5, p. 809, Jul. 2011.

\bibitem{Colton2001}
D.~Colton and B.~D. Sleeman, ``An approximation property of importance in
  inverse scattering theory,'' \emph{Proceedings of the Edinburgh Mathematical
  Society}, vol.~44, no.~3, pp. 449--454, 2001.

\bibitem{Simons2010}
\BIBentryALTinterwordspacing
F.~J. Simons, \emph{Slepian Functions and Their Use in Signal Estimation and
  Spectral Analysis}.\hskip 1em plus 0.5em minus 0.4em\relax Berlin,
  Heidelberg: Springer Berlin Heidelberg, 2010, pp. 891--923. [Online].
  Available: \url{https://doi.org/10.1007/978-3-642-01546-5_30}
\BIBentrySTDinterwordspacing

\bibitem{Slepian1964}
D.~Slepian, ``Prolate spheroidal wave functions, {Fourier} analysis and
  uncertainty—{IV}: extensions to many dimensions; generalized prolate
  spheroidal functions,'' \emph{Bell System Technical Journal}, vol.~43, no.~6,
  pp. 3009--3057, 1964.

\bibitem{Slepian1961}
D.~Slepian and H.~O. Pollak, ``Prolate spheroidal wave functions, {Fourier}
  analysis and uncertainty—{I},'' \emph{Bell System Technical Journal},
  vol.~40, no.~1, pp. 43--63, 1961.

\bibitem{Weck2004}
N.~Weck, ``Approximation by {Herglotz} wave functions,'' \emph{Mathematical
  methods in the applied sciences}, vol.~27, no.~2, pp. 155--162, 2004.

\bibitem{Melenk1999}
J.~M. Melenk, ``Operator adapted spectral element methods {I}: harmonic and
  generalized harmonic polynomials,'' \emph{Numerische Mathematik}, vol.~84,
  pp. 35--69, 1999.

\bibitem{Nehorai1994}
A.~Nehorai and E.~Paldi, ``Acoustic vector-sensor array processing,''
  \emph{IEEE Transactions on signal processing}, vol.~42, no.~9, pp.
  2481--2491, 1994.

\bibitem{Colton2001a}
D.~Colton and R.~Kress, ``On the denseness of {Herglotz} wave functions and
  electromagnetic {Herglotz} pairs in {Sobolev} spaces,'' \emph{Mathematical
  methods in the applied sciences}, vol.~24, no.~16, pp. 1289--1303, 2001.

\bibitem{Ikehata2012}
M.~Ikehata, E.~Niemi, and S.~Siltanen, ``Inverse obstacle scattering with
  limited-aperture data,'' \emph{Inverse problems and imaging}, 2012.

\bibitem{Koyama2019}
S.~Koyama and L.~Daudet, ``Sparse representation of a spatial sound field in a
  reverberant environment,'' \emph{IEEE Journal of Selected Topics in Signal
  Processing}, vol.~13, no.~1, pp. 172--184, 2019.

\bibitem{Mignot2013}
R.~Mignot, G.~Chardon, and L.~Daudet, ``Low frequency interpolation of room
  impulse responses using compressed sensing,'' \emph{IEEE/ACM Transactions on
  Audio, Speech, and Language Processing}, vol.~22, no.~1, pp. 205--216, 2013.

\bibitem{Vembu1961}
S.~Vembu, ``Fourier transformation of the n-dimensional radial delta
  function,'' \emph{The Quarterly Journal of Mathematics}, vol.~12, no.~1, pp.
  165--168, 1961.

\bibitem{Bonami2021}
A.~Bonami, P.~Jaming, and A.~Karoui, ``Non-asymptotic behavior of the spectrum
  of the sinc-kernel operator and related applications,'' \emph{Journal of
  Mathematical Physics}, vol.~62, no.~3, mar 2021.

\end{thebibliography}

\end{document}